\documentclass{kluwer}    
\usepackage{psfig}
\newdisplay{guess}{Conjecture}

\begin{document}
\begin{article}
\begin{opening}
\title{Accretion disks around black holes
  with account of magnetic fields
\thanks{{Partial
funding provided by RFBR grant 02-02-16900, INTAS grant 00491,
and Astronomy Programm "Nonstationary phenomena in
astrophysics"}}}
\author{Gennady \surname{Bisnovatyi-Kogan}}
\runningauthor{G.Bisnovatyi-Kogan}
\runningtitle{Accretion disks around black holes}
\institute{Space Research Institute RAN, Moscow, Russia \\and\\Joint
Institute of Nuclear Researches, Dubna, Russia}
\date{}

\begin{abstract}
Accretion disks are observed in young stars,
cataclysmic variables, binary X-ray sources et al. Accretion disk
theory was first developed as a theory with the local heat
balance, where the whole energy produced by a viscous heating was
emitted to the sides of the disk. Important part
of this theory was the phenomenological treatment of the
turbulent viscosity, known the `` alpha'' prescription, where the
$(r \phi)$ component of the stress tensor was connected with the pressure
as $\alpha P$. Sources of turbulence in the accretion disk are
discussed, including hydrodynamic turbulence, convection and
magnetic field role. Optically thin solution and advective disks
are considered. Related problems of mass ejection from magnetized accretion
disks and jet formation are discussed.
\end{abstract}

\end{opening}

\section{Introduction}

  Accretion is the main source
    of energy in many astrophysical
objects, including different types
of binary stars, binary X-ray sources,
most probably quasars and
active galactic nuclei (AGN).
    Accretion onto stars,
including neutron stars,
terminates at an inner boundary.
   This may be the stellar surface,
or the outer boundary of a magnetosphere
for strongly magnetized stars.
   We may be sure in this case, that all
gravitational energy of the
falling matter will be transformed into heat
and radiated outward.

   The situation is quite different
for sources containing black holes, which are
discovered in some binary X-ray
sources in the galaxy, as well as in many
AGN. Here matter is falling
to the horizon, from where
no radiation arrives,
so all luminosity is formed
on the way to it. A high efficiency of accretion into a black hole
takes place only when matter is magnetized (Schwartzman, 1971), or
has a large angular momentum, when accretion disk is formed.

Intensive development of the accretion disk theory began after
birth of the X-ray astronomy, when luminous X-ray sources in
binary systems have been discovered, in which accretion was the
only possible way of the energy production.
The X-ray astronomy was born after the rocket launch in 1961 in
USA by the group of physicists headed by R. Giacconi. The first and
the brightest X-ray source outside the solar system, Sco X-1, was
discovered during this flight. In subsequent time the main
discoveries in X-ray astronomy have been done from satellites. The
main discovery of the first X-ray satellite UHURU, launched in 1970
had been X-ray pulsars - neutron stars in binary systems. The
fundamental importance has also a discovery of the first real black
hole candidate in the Cyg X-1 binary source. The next X-ray
satellite EINSTEIN, launched in 1978 had a good angular resolution
and sensitivity, and more than 50 000 new sources, mainly
extragalactic, had been discovered there. These two satellites
had been also constructed in the team headed by R. Giacconi.

In subsequent years more than 20 X-ray satellites had been launched,
and the most advanced ones CHANDRA (USA) and NEWTON (ESA) are
operating now.

\section{Standard accretion disk model}

\begin{figure}[ht]
\centerline{
\psfig{file=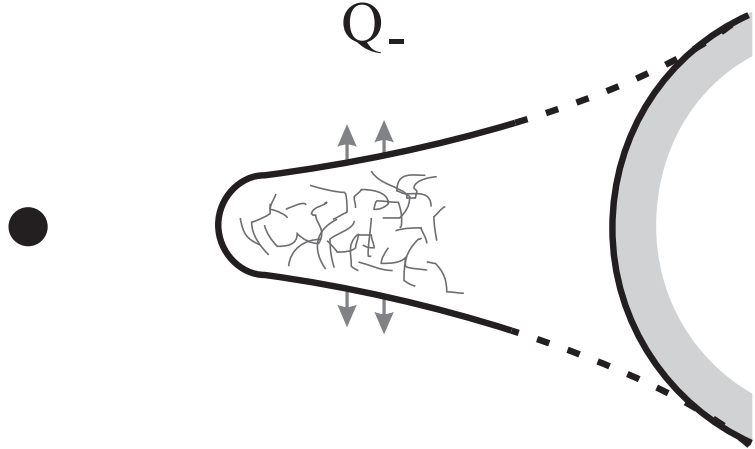,width=10cm,angle=-0}
}
\caption{Schematic picture of a disc accretion into a black hole
in the standard model.}
\label{figu3}
\end{figure}

The small thickness of the disk in
comparison with its radius $h \ll r$
indicate to small importance
of the pressure gradient
$\nabla P$ in comparison with
gravity and inertia forces.
That leads to a simple
radial equilibrium equation
denoting the balance between the last two
forces occuring when the angular
velocity of the disk $\Omega$ is equal to
the Keplerian one $\Omega_K$,

\begin{equation}
\label{ref1.1}
\Omega=\Omega_K=\left(\frac{GM}{r^3}\right)^{1/2}.
\end{equation}
In the ``standard''
accretion disk model the
relation (\ref{ref1.1}) is suggested to be fulfilled
all over the disk, with an inner
boundary at the last stable orbit, $r_{in}=3r_g$,
$r_g=2GM/c^2$ is the gravitational Schwarzschild radius of a
black hole.

The equilibrium equation in the
vertical $z$-direction is determined by a
balance between the gravitational
force and pressure gradient, which
for a thin disk leads to
algebraic one, determining the
half-thickness of the disk as

\begin{equation}
\label{ref1.3}
h \approx \frac{1}{\Omega_K}
\left(2\frac{P}{\rho}\right)^{1/2}.
\end{equation}
The  $\phi$ component of the
Navier-Stokes equation has an integral
in a stationary case which represents the conservation of angular
momentum

\begin{equation}
\label{ref1.4}
\dot M(j-j_{in})=-2\pi r^2\,2ht_{r\phi},\quad t_{r\phi}=
\eta r\frac{d\Omega}{dr}.
\end{equation}
Here $j=v_{\phi}r=\Omega r^2$ is
the specific angular momentum,
$t_{r\phi}$ is the component of the
viscous stress tensor, $\dot M>0$ is a mass
accretion rate, $j_{in}$ is an integration constant.
Multiplication of  $j_{in}$ by $\dot M$,
gives a difference between viscous
and advective flux of the
angular momentum in the disk.
For the accretion into a black hole it is usually assumed,
that on the last stable orbit the
gradient of the angular velocity is
zero, corresponding to zero
viscous momentum flux. In that case

\begin{equation}
\label{ref1.5}
j_{in}=\Omega_K r_{in}^2,
\end{equation}
which is the Keplerian
angular momentum of the matter on the last
stable orbit.

The choice of the viscosity
coefficient is the most difficult and
speculative aspect of the
accterion disk theory.
 Observations
of X-ray binaries had shown, that there should be high viscosity
in the accretion disks.
In the paper of Shakura (1972)
it was suggested, that matter in the disk
is turbulent, and described by
a viscous stress tensor,
parametrized as

\begin{equation}
\label{ref1.7}
t_{r\phi}=-\alpha\rho v_s^2 = -\alpha P,
\end{equation}
where $\alpha$ is a dimensionless constant and $v_s$ is
the sound speed.
This simple parametrization  corresponds to
a turbulent viscosity
coefficient $\eta_t\approx \rho v_t l$
with an average turbulent velocity
$v_t$ and mean free path of the
turbulent element $l$. It follows from
the definition of $
t_{r\phi}$ in (\ref{ref1.4}),
when we take $l \approx h$
from (\ref{ref1.3})

\begin{equation}
\label{ref1.8}
t_{r\phi}=\rho v_t h r \frac{d\Omega}{dr}
\approx \rho v_t v_s =-\alpha
\rho v_s^2,
\end{equation}
where a coefficient $\alpha<1$ relates the turbulent and sound speeds
$v_t=\alpha v_s$.
The presentations of $t_{r\phi}$ in (\ref{ref1.7}) and
(\ref{ref1.8}) are equivalent.
Only when the angular velocity
differs considerably from the
Keplerian is the first relation, on the
right-hand side of (\ref{ref1.8}),  preferable.
That does not appear
(by definition) in the standard
theory, but may happen when advective
terms are included.

Development of a turbulence
in the accretion disk cannot be justified
simply, because a Keplerian disk
is stable in linear approximation to
the development of axially symmetric perturbations,
conserving the angular momentum.
It was suggested by Ya.B.Zeldovich,
that in presence of very large Reynolds number
${\rm Re}=\frac{\rho v l}{\eta}$
the amplitude of perturbations
at which nonlinear effects become important
is very low, so in this situation
turbulence may develop due to
nonlinear instability even when
the disk is stable in linear approximation.
Another source of viscous stresses
may arise from a magnetic field,
but as was indicated by Shakura (1972),
that magnetic stresses cannot
exceed the turbulent ones.

It was shown by Bisnovatyi-Kogan
and Blinnikov (1977), that inner
regions of a highly luminous
accretion discs where pressure is dominated
by radiation, are unstable to
vertical convection. Development of this
convection produce a turbulence,
needed for a high viscosity (and also leads to formation of a
hot corona above this region of the disk). In the colder
regions with incomplete
ionization a behaviour of the accretion disk
becomes more complicated, with
a non-unique solutions, and convective instability
(Cannizzo, Ghosh \& Wheeler, 1982).

For Keplerian angular velocity the angular momentum per unit mass
$j=\omega\,r^2 \sim r^{1/2}$ is growing outside. In this respect it is
similar to the viscid flow between two rotating cylinders (Taylor
experiment), when the inner cylinder is at rest.
Phenomenological analysis of the Taylor experiment, and the onset of
turbulence in the "stable" case of the inner cylinder at rest had been
done by Zeldovich (1981).

There are arguments, both experimental and
theoretical, supporting the hydrodynamic origin of the accretion
disk turbulence. Non-radial perturbations ($\sim e^{im\phi}$)
with a high azimuthal number $m>R/h$ are only slightly influenced
by the rotation, so development of shear instability is possible
for large $m$. The effective length in this case is $l_{eff} \sim
h^2/R$, so the critical $Re_*=\rho v l_{eff}/\eta \approx 10^3$
corresponds to the actual $Re=\rho v h/\eta \approx m Re_*$. In
the Taylor experiment the development of turbulence started at
$Re=10^5$. Analytically a local shear instability in
the stratified accretion disk was found by Richards et al. (2001),
similar results have been obtained earlier (see Glatzel (1991) and
references therein).

Magnetorotational instability of Velikhov (1959)
\& Chandrasekhar (1960) was advocated
by Balbus \& Hawley (1998). Their
numerical experiments which failed to find a
development of hydrodynamic shear instability could not reach the
required $Re \sim 10^5$ in their simulations due to high numerical
viscosity. In real astrophysical objects $Re$ could reach
$10^{10}$, and become even higher.
Results of recent numerical simulations
(Kuznetsov et al., 2004) confirm the idea of
non-linear hydrodynamic instability.

With alpha- prescription
of viscosity the equation of angular
momentum conservation is
written in the plane of the disk as

\begin{equation}
\label{ref1.9}
\dot M(j-j_{in})=
4\pi r^2 \alpha P_0 h.
\end{equation}
When angular velocity is far
from Keplerian the relation
(\ref{ref1.4}) is valid with
a coefficient of a turbulent viscosity

\begin{equation}
\label{ref1.10}
\eta=\frac{2}{3}\alpha\rho_0 v_{s0} h,
\end{equation}
where values with the index ``0''
denotes the plane of the disk.

In the standard theory a heat
balance is local, what means that all
heat produced by viscosity in the
ring between $r$ and $r+dr$ is
radiated through the sides
of disk at the same $r$, see Fig.1.
   The heat production
rate $Q_+$ related to the
surface unit of the disk is written as

\begin{equation}
\label{ref1.11}
Q_+=h\,t_{r\phi}r\frac{d\Omega}{dr}=
\frac{3}{8\pi}\dot M \frac{GM}{r^3}
\left(1-\frac{j_{in}}{j}\right).
\end{equation}
Heat losses by a disk depend on
its optical depth. The first standard disk
model of Shakura (1972) considered
a geometrically thin disk as an optically
thick in a vertical direction.
That implies energy losses $Q_-$ from the disk
due to a radiative conductivity,
after a substitution of
the differential eqiation
of a heat transfer by an algebraic relation

\begin{equation}
\label{ref1.12}
Q_- \approx \frac{4}{3} \frac{acT^4}{\kappa \Sigma}.
\end{equation}
Here $a$ is the usual radiation
energy-density constant,
$c$ is a speed of light,
$T$ is a temperature in the disk plane,
$\kappa$ is a matter opacity,
and a surface density

\begin{equation}
\label{ref1.13}
\Sigma=2\rho h,
\end{equation}
here and below
$\rho,\, T,\,P$ without the
index "0" are related to the disk plane.
The heat balance equation
is represented by a relation

\begin{equation}
\label{ref1.14}
Q_+=Q_-,
\end{equation}
A continuity equation in the
standard model of the stationary accretion flow
is used for finding of a
radial velocity $v_r$

\begin{equation}
\label{ref1.14a}
v_r=\frac{\dot M}{4\pi rh\rho}=
\frac{\dot M}{2\pi r\Sigma}.
\end{equation}
Equations (\ref{ref1.1}),(\ref{ref1.3}),
(\ref{ref1.9}),(\ref{ref1.13}),
(\ref{ref1.14}), completed
by an equation of state $P(\rho,T)$ and relation
for the opacity
$\kappa=\kappa(\rho, T)$
represent a full set of equations
for a standard disk model.
   For power low equations of state of an ideal gas
$P=P_g=\rho {\cal R} T$ (${\cal R}$
is a gas constant), or radiation pressure
$P=P_r=\frac{aT^4}{3}$,
and opacity in the form of electron scattering
$\kappa_e$, or Karammer's formula $\kappa_k$,
the solution of a standard
disk accretion theory is obtained
analytically (Shakura, 1972;
Novikov, Thorne, 1973; Shakura, Sunyaev, 1973).
The structure of the accretion disk around the black hole in the
standard model is represented in Fig.2.

\begin{figure}[ht]
\centerline{
\psfig{file=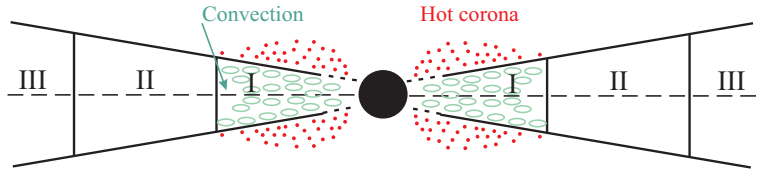,width=10cm,angle=-0}
}
\caption{Sketch of picture of a disk accretion on to a black hole
at sub-critical luminosity, from Bisnovatyi-Kogan (1985).}
\label{fig4}
\end{figure}

\section{Observational evidences of the existence of Black Holes}

\subsection{Observational identification of black holes}

 Black holes (BH) of stellar masses have been observed in the
 galactic binary X-ray sources, and supermassive BH (SBH)
 had been found in the
 nuclei of active galaxies (AGN), having masses $10^7 - 10^9$
 solar masses. In the galactic binary X-ray sources masses have been
 measured on the base of Kepler law: the compact stellar object is
 qualified as BH if its mass exceeds the mass of a stable neutron
 star, about 2.5 solar masses.
 X-ray binaries in our Galaxy
 containing black hole with low mass companion show global
 accretion disk instabilities, observed as soft X-ray transients -
 X-ray novae. The light curves of two X-ray novae in optical and
 X-ray bands are shown in Figs.3,4, see also review of Cherepashchuk
 (2000).
 The mass functions of the compact object in X-ray novae, containing
black holes are 3 -- 6 Solar masses.

\begin{figure}[ht]
\centerline{
\psfig{file=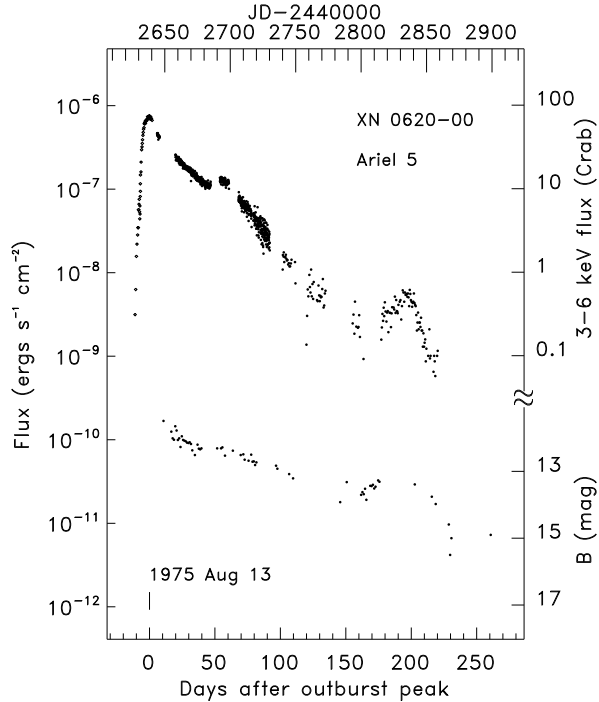,width=8cm,angle=-0}
}
\caption{X-ray and optical light curves of the black-hole LMXBT A0620-00 (from
Chen et al. 1997).}
\label{fig9a}
\end{figure}
\begin{figure}[ht]
\centerline{
\psfig{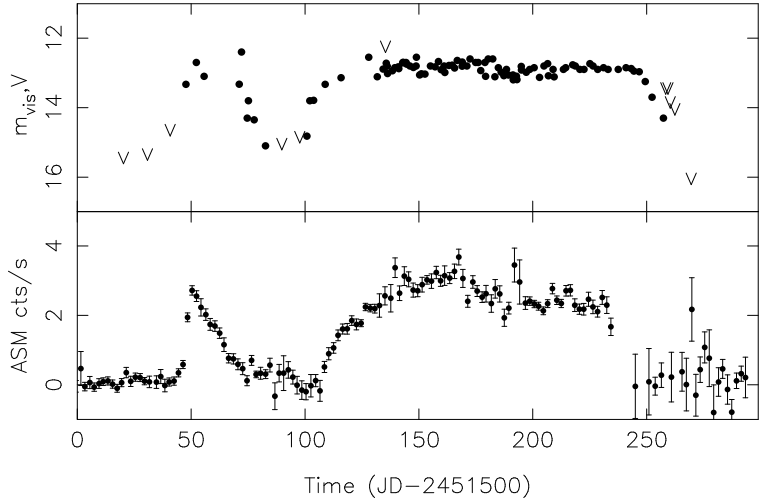}
}
\caption{X-ray and optical light curves of the black-hole LMXBT
 XTE J1118+48 (adapted from Lasota, 2001).
 The `black-hole' classification here requires
confirmation.}
\label{fig9b}
\end{figure}
SBH in AGN are found by optical,
 X-ray and radio observations. Optical observations show strong
 concentration of light to the center, and existence of the
 accretion disk by the distribution of rotational velocity around
 the center. The example of such curve for the galaxy M87
 with the mass of SBH about three billion solar masses is given
 in Fig.5 from Macchetto et al.(1997), see also Ho (1999). X-ray
 observations have revealed an existence of very broad emission
 Fe K$\alpha$ lines in the X-ray spectra of AGN. The width
 corresponding to about one third of the speed of light may
 originate only near the relativistic object. The shape of the
 line is fitted well by the radiation from the accretion disk
 around SBH, which may be described by Schwarzschild or
 Kerr metric. This spectrum given in Fig.6 is representing the composite
 spectrum of Seyfert 1 galactic nuclei, obtained by Nandra et
 al.(1997). The most precise measurements have been done
 by radio VLBI observations of the water maser line from the
 nucleus of the Seyfert galaxy NGC 4258. The Keplerian rotational
 curve is obtained with very high accuracy ($<1\%$), implying the
 mass of 36 millions of solar masses within 0.13 pc. This
 rotational curve is represented in Fig.7 from Ho (1999).

\begin{figure}[ht]
\centerline{
\psfig{file=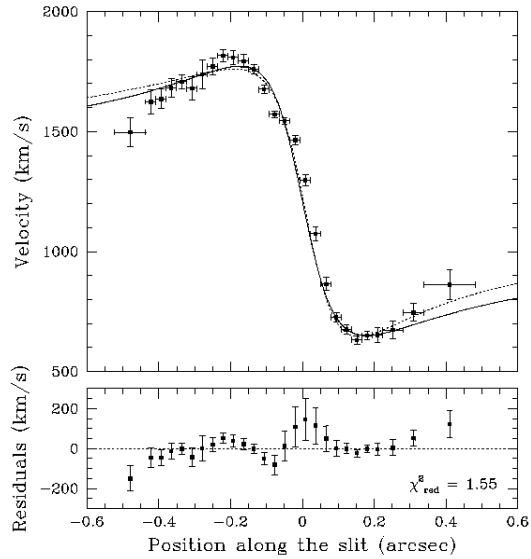,width=8cm,angle=-0}
}
\caption{Optical emission-line rotation curve for the nuclear
disk in M87. The two curves in the upper panel correspond to
Keplerian thin disk models, and the bottom panel shows the
residuals for one of the models, from Macchetto et al. (1999).
}
\label{macchet}
\end{figure}

\begin{figure}[ht]
\centerline{
\psfig{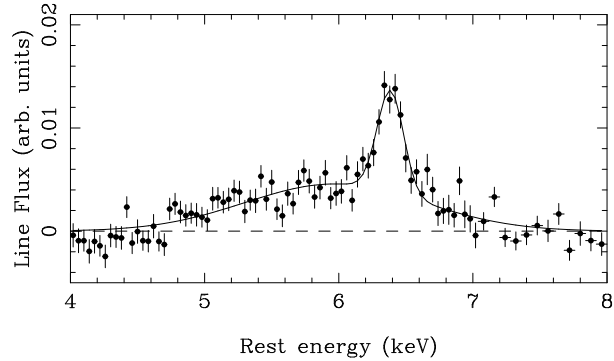}
}
\caption{The Fe K  line in the composite spectrum of Seyfert 1
nuclei. The solid line is a fit to the line profile using two
Gaussians, a narrow component centered at 6.4 keV and a much
broader, redshifted component, from Nandra et al. (1997).
}
\label{nandra}
\end{figure}

\begin{figure}[ht]
\centerline{
\psfig{file=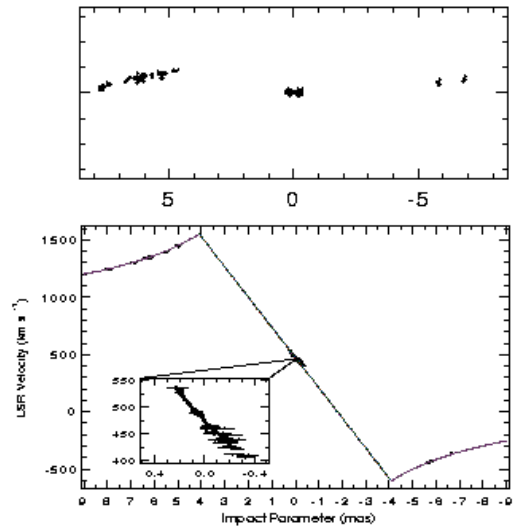,width=8cm,angle=-0}
}
\caption{Water maser emission in NGC 4258 (Miyoshi et al. 1995). Top:
 spatial distribution of the maser features;
bottom: rotation curve. Adapted from
Ho et al. (1999).
}
\label{n4258}
\end{figure}

\subsection{Jets from accretion disks}

\begin{figure}[h]
\centerline{
\psfig{file=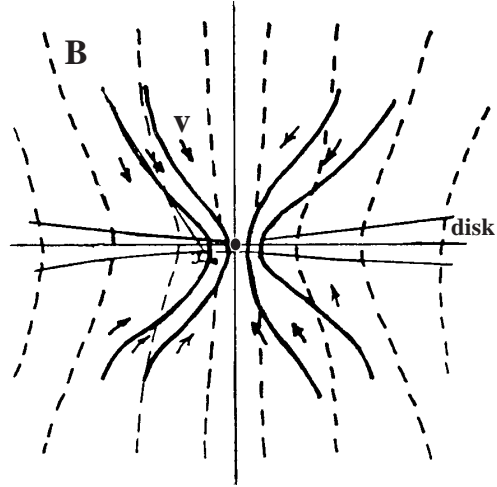,width=8cm,angle=-0}
}
\caption{Sketch of the magnetic field threading an accretion disk shown
increase of the field owing to flux freezing in the accreting disk
matter, from Bisnovatyi-Kogan and Ruzmaikin (1976).}
\label{figbkruz}
\end{figure}

\begin{figure}[h]
\centerline{
\psfig{file=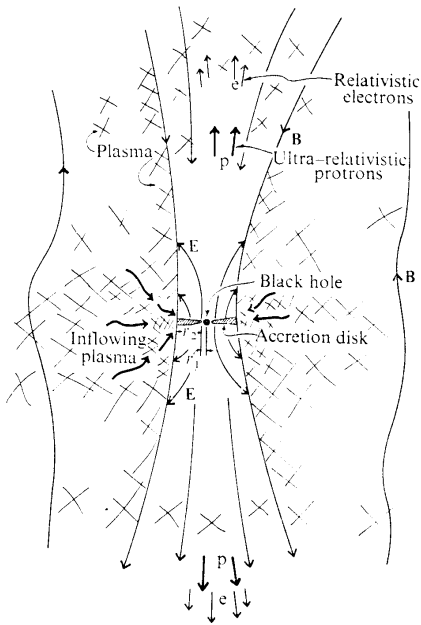,width=6cm,angle=-0}
}
\caption{Sketch of the electromagnetic outflows from the two sides of the disk
owing to the Faraday unipolar dynamo action of a rotating
magnetized disk from Lovelace (1976).}
\label{figlovjet}
\end{figure}

\begin{figure}[h]
\centerline{
\psfig{file=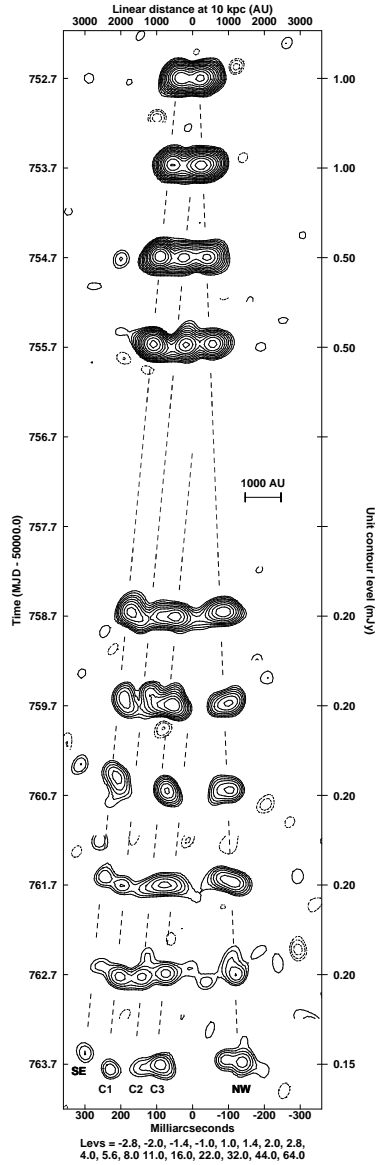,width=5cm,angle=-0}
}
\caption{A sequence of ten epochs of radio imaging of relativistic
ejections from the black hole candidate X-ray binary GRS 1915+105
using MERLIN at 5 GHz.
The figure has been rotated by 52 degrees to form the
montage. For an
estimated distance to the source of 11 kpc the approaching components
have an apparent transverse velocity of $1.5c$. Assuming an intrinsically
symmetric ejection and the standard model for apparent superluminal
motions, an intrinsic bulk velocity for the ejecta is
$0.98^{+0.02}_{-0.05}c$ at an angle to the line of sight of $66 \pm 2$
degrees (at 11 kpc).
The apparent curvature of the jet is probably real, although the cause
of the bending is uncertain.
 (from Fender, 1999).}
\label{jetfenm}
\end{figure}

\begin{figure}[h]
\centerline{
\psfig{file=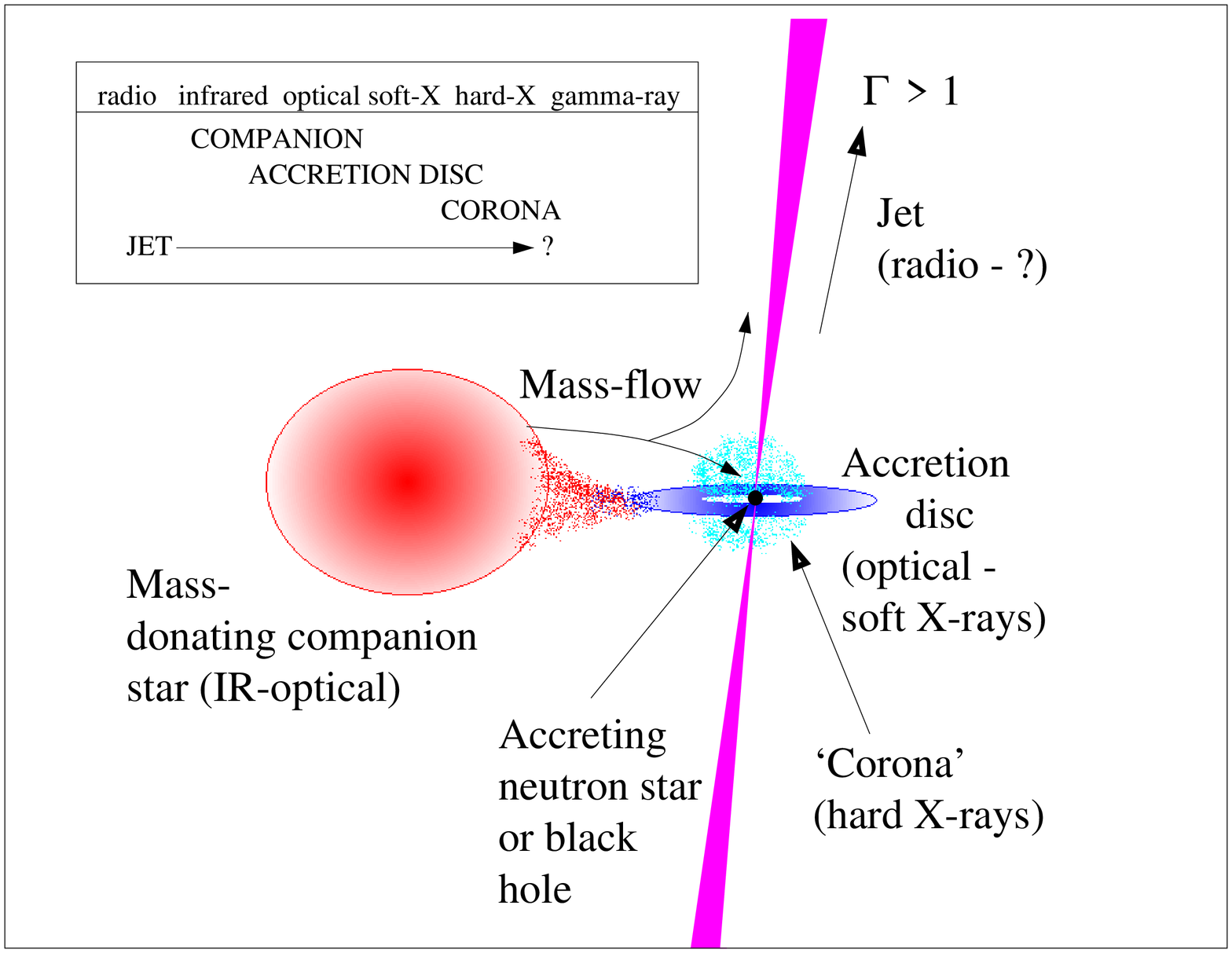,width=7cm,angle=-90}
}
\caption{Schematic picture of jet formation in
binary system
 (from Fender, 2001).}
\label{jetfen}
\end{figure}
   Powerful,
highly-collimated, oppositely directed jets are observed in AGN
and quasars, and in the ``microquasars'' -- old compact stars in
X-ray binaries (Mi\-ra\-bel and Rodriguez, 1994). Highly collimated
emission line jets are seen in young stellar objects.
    Different
ideas and models have been put forward to explain astrophysical
jets (see reviews by Begelman, Blandford and Rees, 1984;
Bisnovatyi-Kogan 1993;
   Lovelace et al. 1999).
 Recent observational and
theoretical work favors models where twisting of an ordered
magnetic field threading an accretion disk
around a black hole acts to magnetically
accelerate the jets as proposed by Lovelace (1976).
   The nature of the ordered magnetic
field threading an accretion disk envisioned by Bisnovatyi-Kogan
and Ruzmaikin (1976) is shown in Fig. 8.
   Fig. 9 shows the outflows from
a disk with such an ordered field from Lovelace (1976).
Jet formation in the microquasar GRS 1915+105 may be seen in
Fig.10 from radio observations of Fender (1999). Picture of the jet
from microquasar XTE J1550-564 obtained from X-ray satellite Chanrda
are given in the paper of Kaaret et al. (2003).
 Formation
of a jet outflowing  from the accretion disk in the X-ray binary
is shown schematically in Fig.11 from Fender (2001).

\begin{figure}[h]
\centerline{
\psfig{file=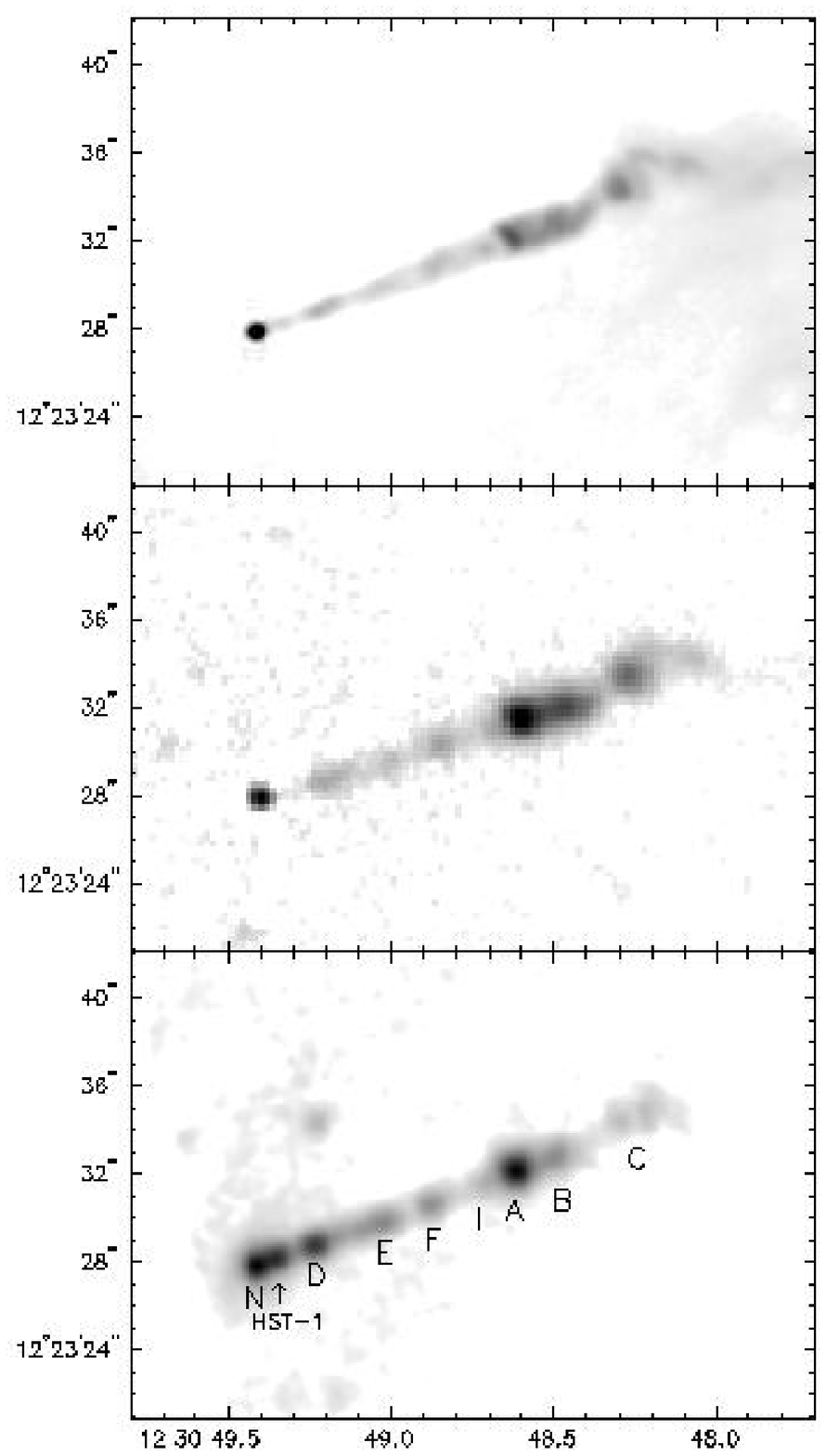,width=8cm,angle=-0}
}
\caption{Grey scale representations of a 6 cm radio (top panel),
an optical V band (middle panel) and the Chandra X-ray (bottom
panel, 0.1 - 10 keV band) image. In the radio image, the grey
scale is proportional to the square root of the brightness, in the
optical image, the grey scale is also proportional to the square
root of the brightness. The labels in the lower panel refer to the
knots vertically above the label. N is the nucleus, from Wilson
and Yang (2002).}
\label{m87}
\end{figure}

\begin{figure}[h]
\centerline{
\psfig{file=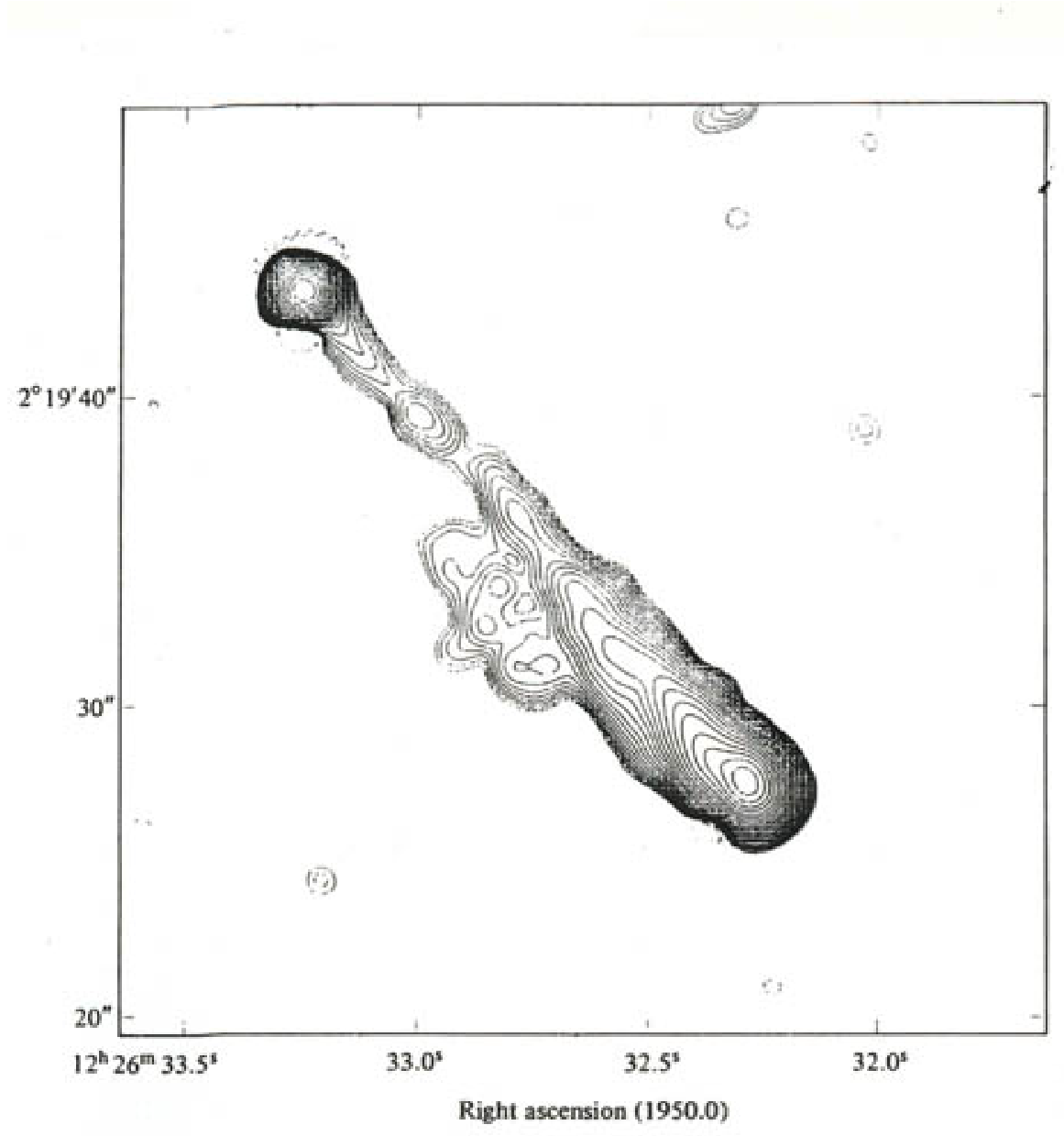,width=8cm,angle=-0}
}
\caption{MERLIN map of 3C 273 at 408 MHz. Resolution: 1.0 arcsec,
from Conway et al. (1993)}
\label{3c273}
\end{figure}

Observations pictures of extragalactic jets, one and two sided,
are very numerous, and are obtained  in several spectral bands
from radio until X-rays. Two most famous jets from AGN are shown
in Fig.12 (M 87, all bands), and Fig.13 (quasar 3C 273, radio).
Impressive pictures in X rays obtained from Chandra
are represented by Marshall et al. (2001) and Harris et al. (2003)
for the jet in M 87; and by Marshall et al. (2000) and Sambruna
et al. (2001) for the jet in the quasar 3C 273. The image of very extended radio jet
from the nucleus of the galaxy IC 4296, with the length about 400 kpc is
represented by Killen et al. (1986).

\section{Farther development of accretion theory}
Few years after appearence of
the standard model it was found that
in addition to the opically
thick disk solution there is another branch
of the solution for the disk
structure with the same input parameters
$M,\,\dot M,\,\alpha$ which
is also self-consistent and has a small
optical thickness
(Shapiro, Lightman, Eardley, 1976). Suggestion of
the small optical thickness
implies another equation of energy losses,
determined by a volume emission

\begin{equation}
\label{ref1.15}
Q_- \approx q\, \rho\,h,
\end{equation}
where due to the Kirchoff law the
emissivity of the unit of a volume $q$ is
connected with a Plankian averaged opacity
$\kappa_p$ by an approximate relation
$q \approx acT_0^4 \kappa_p$.
In the optically thin limit the
pressure is determined by a gas $P=P_g$.
Analytical solutions are obtained
here as well, from the same
equations, with volume losses and gas pressure.
In the optically thin solution
the thickness of the disk is larger then
in the optically thick one, and density is lower.

\begin{figure}[ht]
\centerline{
\psfig{file=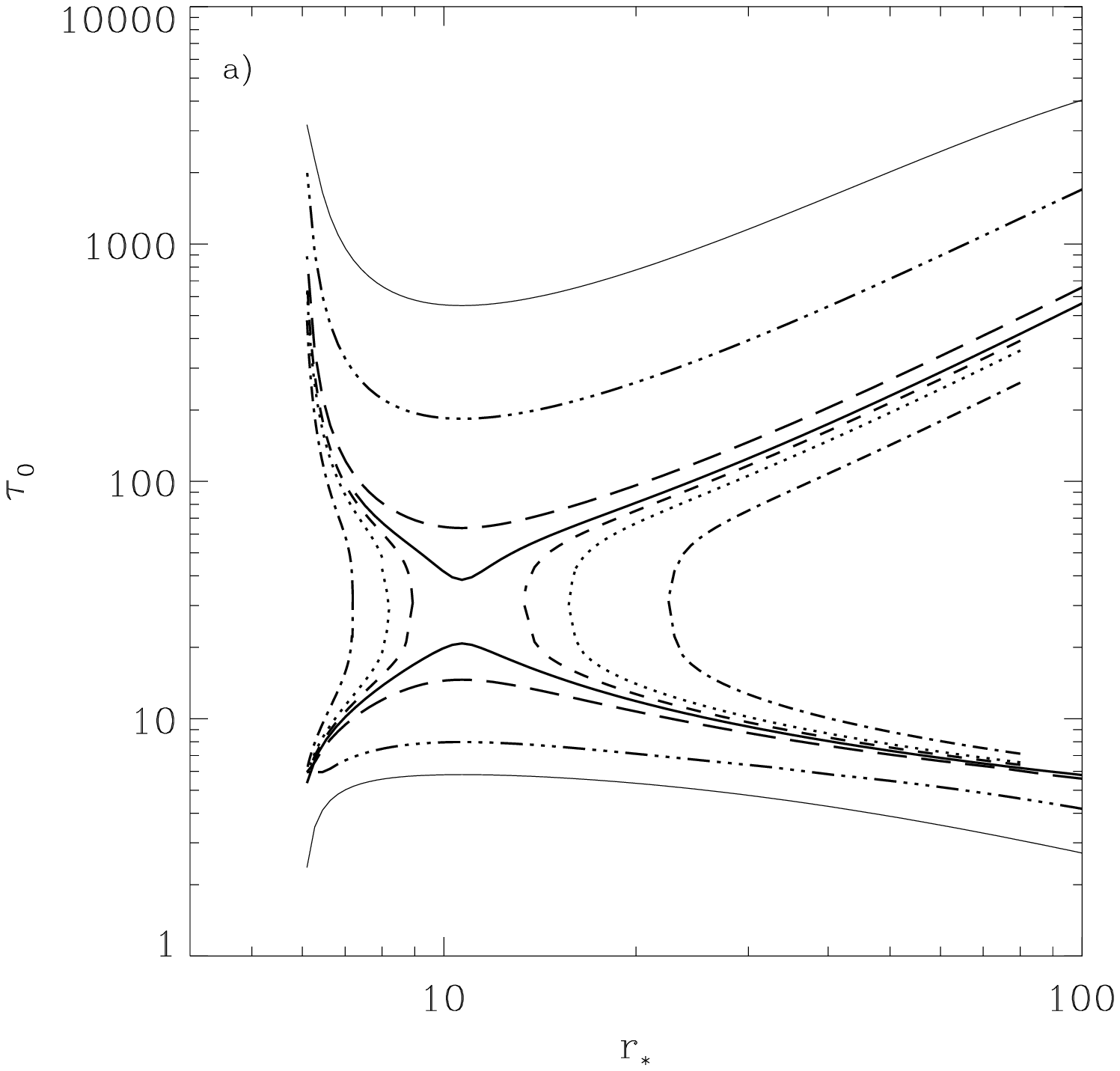,width=7cm,angle=-0}}
\caption{The dependences of the optical depth $\tau_0$ on
radius, $r_*=r/r_{g}$, for the case $M_{BH}=10^8\;M_\odot$,
$\alpha=1.0$ and different values of $\dot m$. The thin solid,
dot-triple dash, long dashed, heavy solid, short dashed, dotted
and dot-dashed curves correspond to $\dot m=1.0, 3.0, 8.0, 9.35,
10.0, 11.0, 15.0$, respectively. The upper curves correspond to
the optically thick family, lower curves correspond to the
optically thin family, from Artemova et al. (1996).}
\label{figartem}
\end{figure}

\begin{figure}[h]
\centerline{
\psfig{file=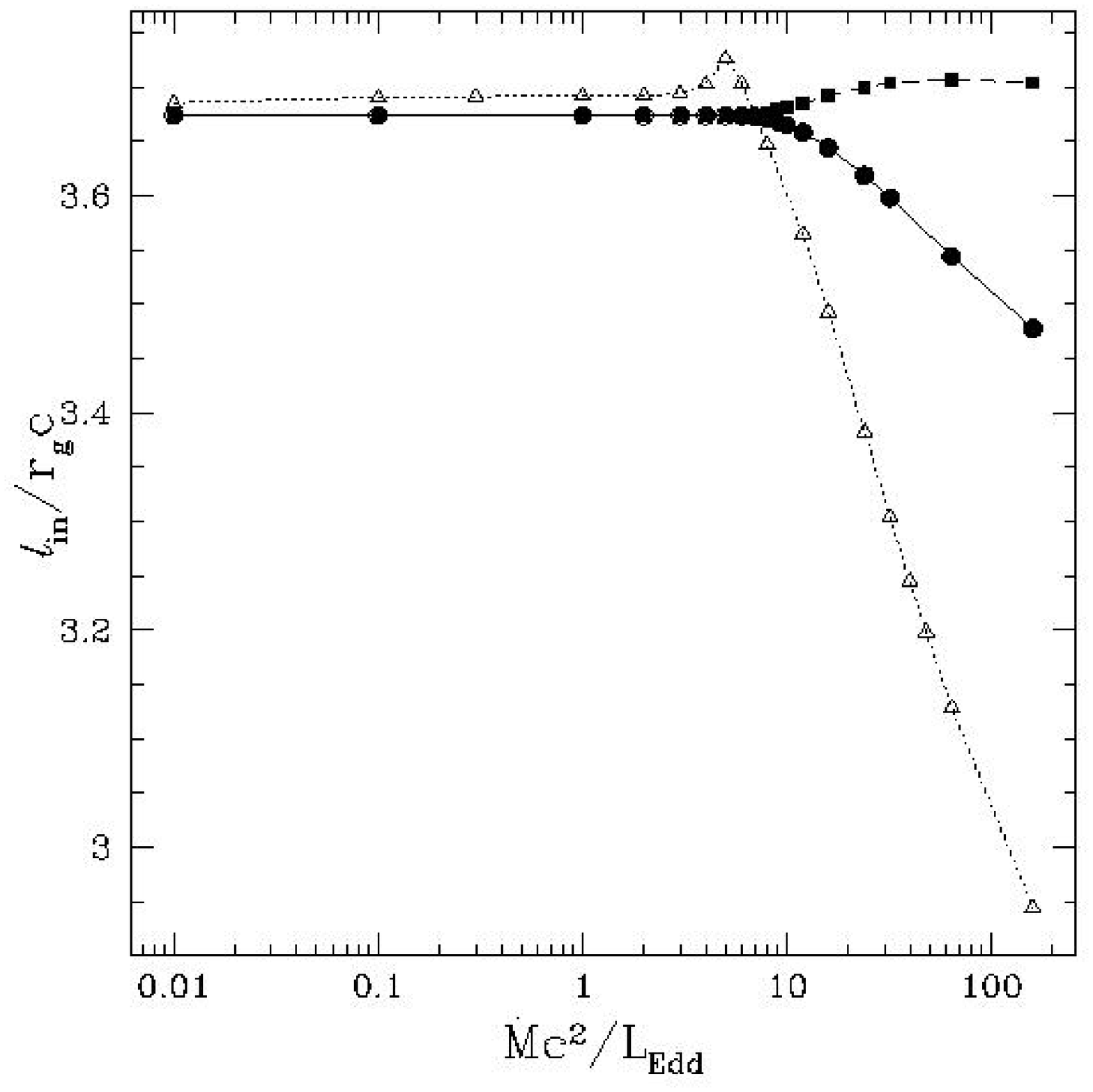,width=7cm,angle=-0}
}
\caption{The specific angular momentum
$j_{\rm in} \equiv {\ell}_{\rm in}$
as a function of the mass accretion rate $\dot{M}$
for different viscosity parameters
$\alpha=0.01$ (squares), $0.1$ (circles) and $0.5$ (triangles),
corresponding to viscosity prescription (\ref{ref1.7}).
The solid dots represent models with the saddle-type
inner singular points, whereas the empty dots
correspond to the nodal-type ones,
from Artemova et al. (2001).}
\label{figfa1}
\end{figure}

\begin{figure}[h]
\centerline{
\psfig{file=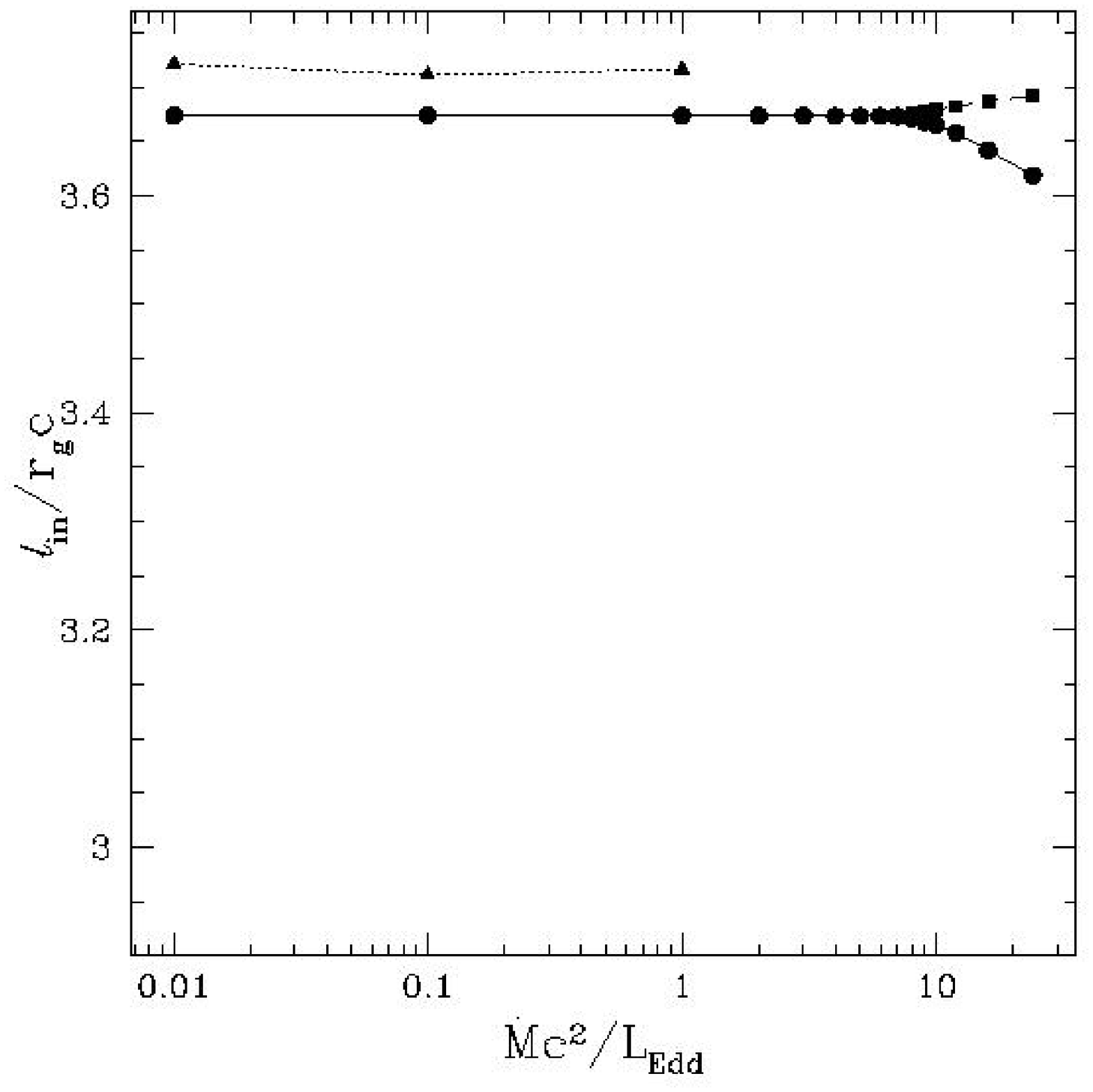,width=7cm,angle=-0}
}
\caption{The specific angular momentum
$j_{\rm in}\equiv {\ell}_{\rm in} $
as a function of the mass accretion rate $\dot{M}$
for different viscosity parameters
$\alpha=0.01$ (squares), $0.1$ (circles) and $0.5$ (triangles),
corresponding to viscosity prescription (\ref{ref1.8}).
The solid dots represent models with the saddle-type
inner singular points,
whereas the empty dots correspond to the nodal-type ones,
from Artemova et al. (2001).}
\label{figfa2}
\end{figure}

\begin{figure}[h]
\centerline{
\psfig{file=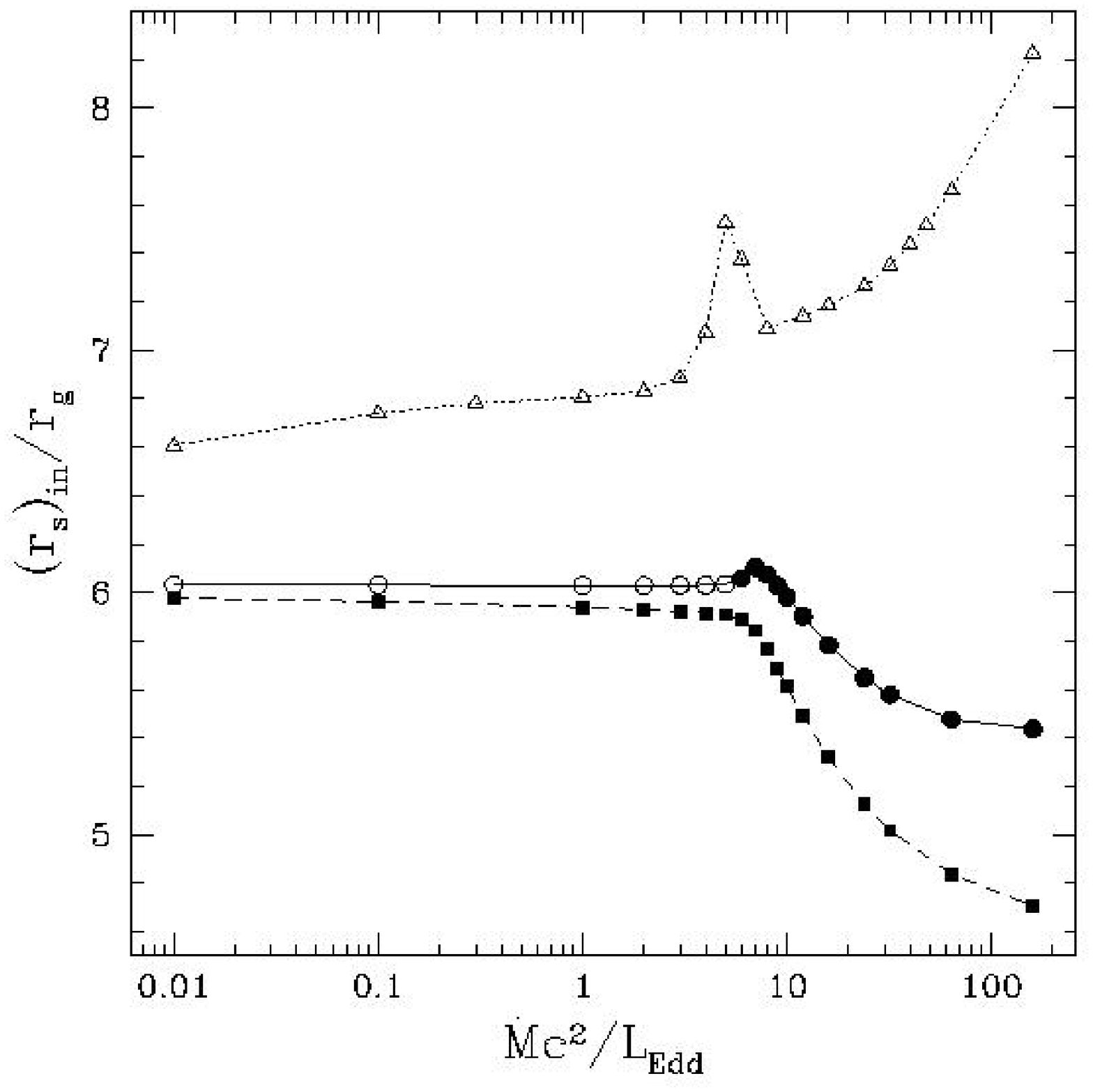,width=7cm,angle=-0}
}
\caption{The position of the inner singular points
as a function of the mass accretion rate $\dot{M}$
for different viscosity parameters
$\alpha=0.01$ (squares), $0.1$ (circles) and $0.5$ (triangles),
corresponding to viscosity prescription (\ref{ref1.7}).
The solid dots represent models with the saddle-type
inner singular points, whereas
the empty dots correspond to the nodal-type ones,
from Artemova et al. (2001).}
\label{figfa3}
\end{figure}

\begin{figure}[h]
\centerline{
\psfig{file=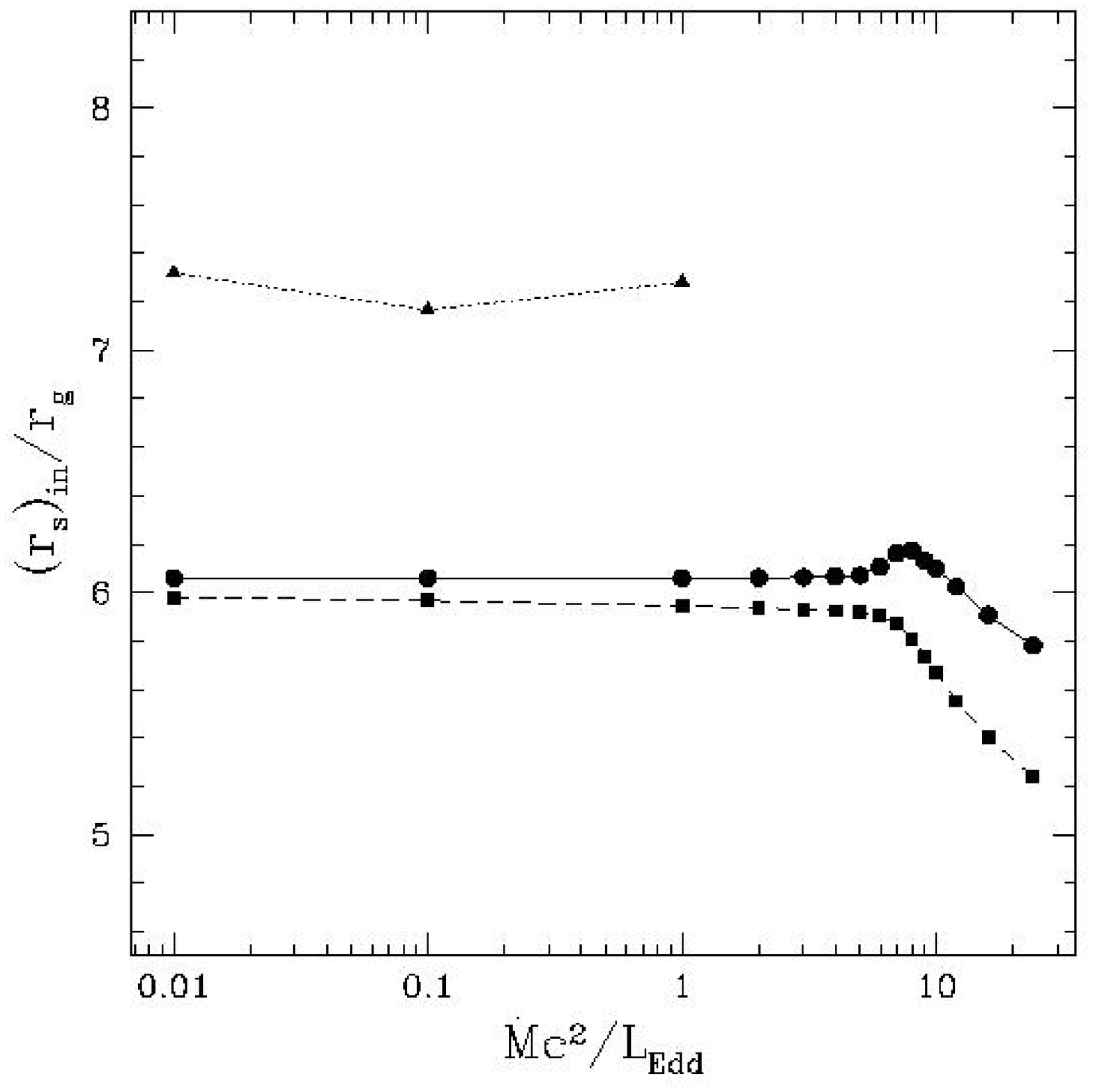,width=7cm,angle=-0}
}
\caption{The position of the inner singular points
as a function of the mass accretion rate $\dot{M}$
for different viscosity parameters
$\alpha=0.01$ (squares), $0.1$ (circles) and $0.5$ (triangles),
corresponding to viscosity prescription (\ref{ref1.8}).
The solid dots represent models with the saddle-type
inner singular points, whereas
the empty dots correspond to the nodal-type ones,
from Artemova et al. (2001).}
\label{figfa4}
\end{figure}
In order to find equations of the disc structure
 valid in both limiting cases of optically thick and optically thin disc,
and smoothly describing transition between them, Eddington
approximation had been used for obtaining formulae for a heat flux
and for a radiation pressure (Artemoma et al., 1996). The
following expressions had been obtained for the vertical energy
flux from the disc $F_0$, and the radiation pressure in the
symmetry plane

\begin{equation}
\label{ref11.15}
 F_{0}={2acT_0^4 \over 3\tau_{0}}\left(1+{4 \over 3\tau_{0}}+
 {2 \over 3\tau_{*}^2}\right)^{-1},\quad
 P_{rad,0}={aT_0^4 \over 3}
 {1+{4 \over 3 \tau_{0}}\over 1+{4 \over 3\tau_{0}}+
 {2\over 3\tau_{*}^2}},
\end{equation}
where
 $ \tau_{0}=\kappa_e \rho h$,
  $\tau_{*}=\left(\tau_{0}\tau_{\alpha 0}\right)^{1/2}$,
  $\tau_{\alpha 0} \approx \kappa_p \rho h$.
At $\tau_0 \gg \tau_* \gg 1$ we have (\ref{ref1.12}) from
(\ref{ref11.15}). In the optically thin limit $\tau_* \ll \tau_0
\ll 1$ we get

\begin{equation}
\label{ref11.17}
 F_{0}=acT_0^4 \tau_{\alpha 0}, \quad
 P_{rad,0}={2 \over 3}acT_0^4 \tau_{\alpha 0}.
\end{equation}
Using $F_0$ instead of $Q_-$ and equation of state $P=\rho {\cal
R} T+P_{rad,0}$, the equations of accretion disc structure
together with equation $Q_+=F_0$, with $Q_+$ from (\ref{ref1.11}),
have been solved numerically by Artemova et al. (1996). It occurs
that two solutions, optically thick and optically thin, exist
separately when luminosity is not very large. Two solutions
intersect at $\dot m=\dot m_b$ and there is no global solution for
accretion disc at $\dot m > \dot m_b$ (see Fig.14). It was
concluded by Artemova et al (1996), that in order to obtain a
global physically meaningful solution at $\dot m > \dot m_b$,
account of advection is needed.

It is clear from physical ground, that when a
local heat production due to viscosity goes to zero near the
inner edge of the disk, the heat
brought by radial motion of matter along the accretion disc
becomes more important. In presence of this advective heating (or
cooling term, depending on the radial entropy $S$ gradient)
written as

\begin{equation}
\label{ref3.1}
Q_{adv}=\frac{\dot M}{2\pi r}T \frac{dS}{dr},
\end{equation}
the equation of a heat balance is modified to $Q_+ + Q_{adv}=Q_-$.
In order to describe self-consistently the structure of the
accertion disc we should also modify the radial disc equilibrium,
including pressure and inertia terms.
Appearance of inertia term leads to transonic radial flow with a
singular point. Conditions of a continuous passing of the solution
through a critical point choose a unique value of the integration
constant $j_{in}$. In the standard local theory $j_{in}=j_{in}^{(0)}$
corresponds to
the keplerian angular momentum on the last stable orbit at $r=3r_g$,
$r_g=\frac{2GM}{c^2}$.
To model the effects of general relativity the
gravitational potential of Paczynski \& Wiita (1980) $\Phi=\frac{GM}{r-r_g}$
had been used in calculations of Artemova et al. (2001). For this potential

\begin{equation}
\label{ref3.2}
j_{in}^{(0)}\equiv l_{in}^{(0)}=\frac{3}{2}\sqrt{\frac{3}{2}} r_g c,
\end{equation}
First approximate solution for the advective
disc structure have been obtained by Paczynski and
Bisnovatyi-Kogan (1981). Accretion disk models with advection in
the optically thick limit have been constructed numerically by
Abramovicz et al. (1988), and improved by Artemova et al. (2001).
Add dynamical and radial pressure gradient term to the equation of radial
equilibrium. Instead of Keplerian angular velocity we obtain radial
hydrodynamic equation. Radial accretion flux becomes supersonic in the
vicinity of the inner last stable orbit. The position of the critical
radius is a proper value of the problem, differs noticeable from the
radius of the last stable orbit at luminosity approaching the critical
Eddington one  $L_{Edd}=\frac{4 \pi cGM}{\kappa}$
Relaxation method
corrected for the existence of critical points had been used in
calculations of Artemova et al. (2001), permitting to find solutions
at large luminosity, formally exceeding the critical Eddington one.
The results of these calculations are represented in Figs. 15-18.

\subsection{Influence of the small-scale magnetic field on the accretion}

While heating
by viscosity is determined mainly
by  ions, and cooling is determined
by electrons, the rate of the energy
exchange between them is important for
a thermal structure of the disk.
The energy balance equations are written
separately for ions and electrons.
For small accretion rates and lower
matter density the rate of energy
exchange due to binary collisions is
so slow, that in the thermal
balance the ions are much hotter then the
electrons.
That also implies a high
disk thickness.

It was
noticed by Narayan and Yu (1995), that advection in this case is becoming
extremely important. It may carry the main energy flux into a black
hole, leaving rather low efficiency of the accretion up to $10^{-4}\,-\,
10^{-5}$ (advective dominated accretion flows - ADAF).
This conclusion is valid only when the effects, connected
with magnetic field annihilation and heating of matter due to it are
neglected.
 To support the condition of equipartition during accretion a
continuous magnetic field reconnection is necessary, leading to annihilation
of the magnetic flux and heating of matter due to Ohmic heating.
The heating of electrons during reconnection is equal or larger
than ion heating. While all electron energy is emitted by magneto-
bremstrahlung radiation, the efficiency of accretion cannot become
less than 0.25 of its standard value (0.06 for Schwarzschild
metrics) (see Bisnovatyi-Kogan and Lovelace, 2001).
In addition, in the highly
turbulent plasma the energy exchange
between ions and electrons may
be strongly enhanced due to presence
of fluctuating electrical fields,
where electrons and ions gain the
same energy. In such conditions
difference of temperatures between ions and
electrons may be negligible.

While heating
by viscosity is determined mainly
by  ions, and cooling is determined
by electrons, the rate of the energy
exchange between them is important for
a thermal structure of the disk.
The energy balance equations are written
separately for ions and electrons.
For small accretion rates and lower
matter density the rate of energy
exchange due to binary collisions is
so slow, that in the thermal
balance the ions are much hotter then the
electrons.
That also implies a high
disk thickness.

It was
noticed by Narayan and Yu (1995), that advection in this case is becoming
extremely important. It may carry the main energy flux into a black
hole, leaving rather low efficiency of the accretion up to $10^{-4}\,-\,
10^{-5}$ (advective dominated accretion flows - ADAF).
This conclusion is valid only when the effects, connected
with magnetic field annihilation and heating of matter due to it are
neglected.
 To support the condition of equipartition during accretion a
continuous magnetic field reconnection is necessary, leading to annihilation
of the magnetic flux and heating of matter due to Ohmic heating
Bisnovatyi-Kogan and Ruzmaikin (1974).
The heating of electrons during reconnection is equal or larger
than ion heating. While all electron energy is emitted by magneto-
bremstrahlung radiation, the efficiency of accretion cannot become
less than 0.25 of its standard value (0.06 for Schwarzschild
metrics) (see Bisnovatyi-Kogan and Lovelace, 2001).
In addition, in the highly
turbulent plasma the energy exchange
between ions and electrons may
be strongly enhanced due to presence
of fluctuating electrical fields,
where electrons and ions gain the
same energy. In such conditions
difference of temperatures between ions and
electrons may be negligible.

\section{Summary}

1. Accretion into black holes (neutron stars) is a main source of
energy in galactic X-ray sources and AGNs.

2. Jets, observed in AGNs and microquasars my be explained by
interaction of the large-scale magnetic field with the accretion disk
in self-consistent picture.

3. Accretion model deviate substantially from the standard (local)
model at high accretion rates (luminosities), where advection effects
are important.

4. Magnetic field action prevent formation of low-efficient flow (ADAF)

5. Turbulence in accretion disks, and turbulent viscosity is created
mainly due to development of non-linear hydrodynamic instabilities
at large Reynolds number.

\acknowledgements
Author is grateful to the conference organizers
for support and hospitality.

\end{article}

\begin{thebibliography}{}

\bibitem[\protect\citeauthoryear{Abramovich et al.}{1988}]{abr88}
Abramovicz, M.~A., B. Czerny, J.~P. Lasota,  and E. Szuszkiewicz.
\newblock{Slim Accretion Disks}
\newblock {\em Astrophys. J.}, 332:646--658,1988.

\bibitem[\protect\citeauthoryear{Artemova et al.}{1996}]{ar1}
Artemova, I.~V., G.~S. Bisnovatyi-Kogan, G.
Bj\"ornsson  and  I.~D. Novikov.
\newblock {Structure of Accretion Disks with Optically
Thick--Optically Thin Transitions}.
\newblock {\em Astrophys. J.}, 456:119--123, 1996.

\bibitem[\protect\citeauthoryear{Artemova et al.}{2001}]{ar01}
Artemova, I.~V., G.~S. Bisnovatyi-Kogan, I.~V. Igumenshchev
and I.~D. Novikov.
\newblock {On the Structure of Advective Accretion Disks at High Luminosity}.
\newblock {\em Astrophys. J.}, 549:1050--1061, 2001.

\bibitem[\protect\citeauthoryear{Balbus and Hawley}{1998}]{bh}
Balbus S.~A. and J.~F. Hawley.
\newblock {Instability, Turbulence, and Enhanced Transport in Accretion Disks}.
\newblock {\em Rev. Mod. Phys.}, 70:1--53, 1998.

\bibitem[\protect\citeauthoryear{Begelman et al.}{1984}]{bb84}  Begelman M.C.,
Blandford R.~D. and M.~J. Rees.
\newblock {Theory of Extragalactic Radio Sources}.
\newblock {\em Rev. Mod. Phys.}, 56:255--351, 1984.

\bibitem[\protect\citeauthoryear{Bisnovatyi-Kogan}{1985}]{bk85}
Bisnovatyi-Kogan, G.~S.
\newblock {X ray Sources in Close Binary Systems: Theoretical Aspects}.
\newblock {\em Bulletin Abastumani Astrophys. Obs.}, No. 58:175--210, 1985.

\bibitem[\protect\citeauthoryear{Bisnovatyi-Kogan}{1993b}]{bkj93}
Bisnovatyi-Kogan, G.`S.
\newblock {Mechanisms of Jet Formation}.
\newblock In  L.~Errico and A.~A.~Vittone, editors,
{\em Proceedings of Int. Conf. Stellar Jets and Bipolar Outflows}, 369--381, 1993.
 Kluwer, Dordrecht.

\bibitem[\protect\citeauthoryear{Bisnovatyi-Kogan and Blinnikov}{1977}]{bkb1}
Bisnovatyi-Kogan G.~S. and S.~I.~Blinnikov.
\newblock {Disk Accretion onto a Black Hole at Subcritical Luminosity}.
\newblock {\em Astron. Ap.}, 59:111--125, 1977.

\bibitem[\protect\citeauthoryear{Bisnovatyi-Kogan and Lovelace}{2001}]{bkl2}
Bisnovatyi-Kogan G.~S. and R.~V.~L. Lovelace.
\newblock {Advective Accretion Disks and Related Problems Including Magnetic Fields}.
\newblock {\em New Astronomy Reviews}, 45:663--742, 2001.

\bibitem[\protect\citeauthoryear{Bisnovatyi-Kogan and Ruzmaikin}{1974}]{bkr74}
Bisnovatyi-Kogan, G.~S. and A.~A. Ruzmaikin.
\newblock {The Accretion of Matter by a Collapsing Star
in the Presence of a Magnetic Field}.
\newblock {\em Astrophys. and Space Sci.}, 28:45--59, 1974.

\bibitem[\protect\citeauthoryear{Bisnovatyi-Kogan and Ruzmaikin}{1976}]{bkr76}
Bisnovatyi-Kogan, G.~S. and A.~A. Ruzmaikin.
\newblock {The Accretion of Matter by a Collapsing Star in the Presence of a
 Magnetic Field. II - Selfconsistent Stationary Picture}.
\newblock {\em Astrophys. and Space Sci.}, 42:401--424, 1976.

\bibitem[\protect\citeauthoryear{Cannizzo et al.}{1982}]{cgw}
Cannizzo, J., P. Ghosh and J.~C. Wheeler
\newblock {Convective Accretion Disks and the Onset of Dwarf Nova Outbursts}.
\newblock {\em Astrophys. J. Lett.}, 260:L83--L86, 1982.

\bibitem[\protect\citeauthoryear{Chandrasekhar}{1961}]{chan}
Chandrasekhar, S.
\newblock {\em Hydrodynamic and Hydromagnetic Stability}.
International Series of Monographs on Physics, Oxford: Clarendon, 1961.

\bibitem[\protect\citeauthoryear{Chen et al.}{1997}]{chen97}
Chen, W., C.~R. Shrader, and M. Livio.
\newblock {The Properties of X-Ray and Optical Light Curves of X-Ray Novae}.
\newblock {\em Astrophys. J.}, 491:312--338, 1997.

\bibitem[\protect\citeauthoryear{Cherepashchuk}{2000}]{cher1}
Cherepashchuk, A.~M.
\newblock {X-ray Nova Binary Systems}.
\newblock {\em Space Sci. Rev.}, 93:473--580, 2000.

\bibitem[\protect\citeauthoryear{Conway et al.}{1993}]{con}
Conway, R. G., S.~T. Garrington,  R.~A. Perley, and J.~A. Biretta.
\newblock {Synchrotron Radiation from the Jet of 3C 273.
II - The Radio Structure and Polarization}.
\newblock {\em Astron. Ap.}, 267:347--362, 1993.

\bibitem[\protect\citeauthoryear{Fender}{1999}]{fen}
Fender, R.
\newblock {Relativistic Jets from X-ray Binaries}.
\newblock {\em Astro-ph/9907050}, 1999.

\bibitem[\protect\citeauthoryear{Fender}{2001}]{fen01}
Fender, R.
\newblock {Relativistic Outflows from X-ray Binaries}.
\newblock {\em Astro-ph/0109502}, 2001.

\bibitem[\protect\citeauthoryear{Glatzel}{1991}]{gla}
Glatzel W.,
\newblock {Instabilities in Astrophysical Shear Flows}.
\newblock {\em Rev. in Modern Astron.}, 4:104--116, 1991.

\bibitem[\protect\citeauthoryear{Harris et al.}{2003}]{har}
Harris, D.~E., J.~A. Biretta, W. Junor, E.~S. Perlman, W.~B. Sparks
and A.~S. Wilson.
\newblock {Flaring X-ray Emission from HST-1, a Knot in the M87 Jet}.
\newblock {\em Astrophys. J.}, 586:L41-L44, 2003.

\bibitem[\protect\citeauthoryear{Ho}{1999}]{ho}
Ho, L.
\newblock {Supermassive Black Holes in Galactic Nuclei:
Observational Evidence and Astrophysical Consequences}.
\newblock {\em Observational Evidence for the Black Holes in the Universe,
Proc. Conference held in Calcutta, January 11-17th, 1998.}, 157--186,
Kluwer, 1999.

\bibitem[\protect\citeauthoryear{Kaaret et al.}{2003}]{kaar}
 Kaaret, P., S. Corbel, J.~A. Tomsick, R.~P. Fender, J.~M. Miller,
J~.A. Orosz, A.~K. Tzioumis and R. Wijnands.
\newblock {X-Ray Emission from the Jets of XTE J1550-564}.
\newblock {\em Astrophys. J.},  582:945--953, 2003.

\bibitem[\protect\citeauthoryear{Killen et al.}{1986}]{kil}
 Killeen, N.~E.~B., G.~V. Bicknell and  R. D.Ekers.
\newblock { The radio galaxy IC 4296 (PKS 1333 - 33).
    I - Multifrequency Very Large Array observations}.
\newblock {\em Astrophys. J.}, 302;306--336, 1986.

\bibitem[\protect\citeauthoryear{Kuznetsov et al.}{2004}]{kuz}
Kuznetsov, O.~A., G.~S.Bisnovatyi-Kogan and D. Molteni.
\newblock {On the development of hydrodynamical turbulence in
accretion discs: numerical simulations}.
\newblock {\em Month. Not. R.A.S.}, submitted, 2004.

\bibitem[\protect\citeauthoryear{Lasota}{2001}]{las01}
Lasota, J.-P.
\newblock {The disc instability model of dwarf novae and
low-mass X-ray binary transients}.
\newblock {\em New Astronomy Reviews}, 45:449--508, 2001.

\bibitem[\protect\citeauthoryear{Lovelace}{1976}]{lov}
Lovelace, R.~V.~E.
\newblock {Dynamo model of double radio sources}.
\newblock {\em  Nature}, 262:649--652, 1976.

\bibitem[\protect\citeauthoryear{Artemova et al.}{1999}]{luk}
Lovelace, R.~V.~E., G.~S. Ustyugova and A.~V. Koldoba.
\newblock {Magnetohydrodynamic Origin of Jets from Accretion Disks}.
\newblock {\em Active Galactic Nuclei and Related Phenomena, IAU Symposium 194,
eds. Y. Terzian, D. Weedman and E. Khachikian (Astron. Soc. of
Pacific: San Francisco)}, 208--217, 1999.

\bibitem[\protect\citeauthoryear{Macchetto et al.}{1997}]{mac}
Macchetto, F.,   A. Marconi, D.~J. Axon, A. Capetti,  W. Sparks
and P. Crane.
\newblock {The Supermassive Black Hole of M87 and the Kinematics
of Its Associated Gaseous Disk}.
\newblock {\em Astrophys. J.},  489:579--600, 1997.

\bibitem[\protect\citeauthoryear{Marshall et al.}{2001}]{mar}
 Marshall, H.~L., D.~E. Harris, J.~P. Grimes, J.~J. Drake
A. Fruscione, M. Juda, R.~P. Kraft, S. Mathur,
S.~S. Murray, P.~M. Ogle, D.~O. Pease, D.~A. Schwartz,
A.~L. Siemiginowska, S.~D. Vrtilek and B.~J. Wargelin.
\newblock {Structure of the X-Ray Emission from the Jet of 3C 273}.
\newblock {\em Astrophys. J.}, 549:L167--L171, 2001.

\bibitem[\protect\citeauthoryear{Marshall et al.}{2002}]{mar1}
Marshall, H.~L., B.~P. Miller, D.~S. Davis, E.~S. Perlman,
M. Wise, C.~R. Canizares and D.~E. Harris.
\newblock {A High-Resolution X-Ray Image of the Jet in M87}.
\newblock {\em Astrophys. J.},  564:683--687, 2002.

\bibitem[\protect\citeauthoryear{Mirabel and Rodriguez}{1994}]{mr94}
Mirabel, I.~F. and L.~F. Rodriguez.
\newblock {A Superluminal Source in the Galaxy}.
\newblock {\em Nature}, 371:46--48, 1994.

\bibitem[\protect\citeauthoryear{Miyoshi et al.}{Miyoshi}]{miy}
Miyoshi, M., J. Moran, J. Herrnstein, L. Greenhill, N. Nakai,
P. Diamond, and M. Inoue.
\newblock {Evidence for a Black Hole from High Rotation
Velocities in a Sub-parsec Region of NGC4258}.
\newblock {\em Nature}, 373:127--129, 1995.

\bibitem[\protect\citeauthoryear{Nandra et al.}{1997}]{nan}
Nandra, K., I.~M. George, R.~F. Mushotzky, T.~J. Turner, T. Yaqoob.
\newblock {ASCA Observations of Seyfert 1 Galaxies.
II. Relativistic Iron K alpha Emission}.
\newblock {\em Astrophys. J.}, 477:602--622, 1997.

\bibitem[\protect\citeauthoryear{Narayan and Yi}{1995}]{nay}
Narayan, R. and I. Yi.
\newblock {Advection-dominated Accretion:
Underfed Black Holes and Neutron Stars}.
\newblock {\em Astrophys. J.}, 452:710--735, 1995.

\bibitem[\protect\citeauthoryear{Novikov and Thorne}{1973}]{nt}
Novikov, I.~D. and K.~S. Thorne.
\newblock {Astrophysics of Black Holes}.
\newblock {\em Black Holes eds. C.DeWitt \& B.DeWitt
(New York: Gordon \& Breach)}, 345--405, 1973.

\bibitem[\protect\citeauthoryear{Paczy\'nski and Bisnovatyi-Kogan}{1981}]{pbk}
Paczy\'nski, B. and G.~S. Bisnovatyi-Kogan.
\newblock {A Model of a Thin Accretion Disk around a Black Hole}.
\newblock {\em Acta Astron.}, 31:283--291, 1981.

\bibitem[\protect\citeauthoryear{Richard et al.}{2001}]{rich}
Richard, D., F. Hersant, O. Dauchot, F. Daviaud, B. Dubrulle and J-P. Zahn.
\newblock {A powerful local shear instability in stratified disks}.
\newblock {\em Astro-ph/0110056}, 2001.

\bibitem[\protect\citeauthoryear{Samburina et al.}{2001}]{sam}
Sambruna, R.~M., C.~M. Urry, F. Tavecchio, L. Maraschi,
R. Scarpa,  G. Chartas and T. Muxlow.
\newblock {Chandra Observations of the X-Ray Jet of 3C 273}.
\newblock {\em Astrophys. J.}, 549:L161-L165, 2001.

\bibitem[\protect\citeauthoryear{Schwartsman}{1971}]{sch71}
Schwartsman, V.~F.
\newblock {Halos around "Black Holes"}.
\newblock {\em  Soviet Astron.},  15:377--388, 1971.

\bibitem[\protect\citeauthoryear{Shakura}{1972}]{sha}
Shakura, N.~I.
\newblock {Disk Model of Gas Accretion on a Relativistic Star in a
Close Binary System}.
\newblock {\em Astron. Zh.}, 49:921--930, 1972
(1973, Sov. Astron. 16, 756).

\bibitem[\protect\citeauthoryear{Shakura and Sunyaev}{1973}]{shs}
Shakura, N.~I. and R.~A. Sunyaev.
\newblock {Black holes in binary systems. Observational appearance}.
\newblock {\em Astron. Ap.},  24:337--355, 1973.

\bibitem[\protect\citeauthoryear{Shapiro et al.}{1976}]{sle}
Shapiro, S.~L., A.~P. Lightman and D.~M. Eardley.
\newblock {A two-temperature accretion disk model for
Cygnus X-1 - Structure and spectrum}.
\newblock {\em Astrophys. J.}, 204:187--199, 1976.

\bibitem[\protect\citeauthoryear{Velikhov}{1959}]{vel}
Velikhov, E.P.
\newblock {Stability of an ideally conducting
liquid flowing between rotating cylinders in a
magnetic field}.
\newblock {\em J. Exper. Theor. Phys.}, 36:1398--1404, 1959.

\bibitem[\protect\citeauthoryear{Zeldovich}{1981}]{zel}
Zeldovich, Ya.~B.
\newblock {On the friction of fluids between rotating cylinders}.
\newblock {\em  Proc. R. Soc. Lond., A }, 374:299--312, 1981.

\end{thebibliography}
\end{document}